\documentclass[printer]{aa}
\usepackage{graphicx}
\usepackage{epsfig}
\usepackage{lscape}
\usepackage{rotating}
\usepackage{natbib}
\usepackage{tabularx}
\usepackage{subfig}
\bibpunct{(}{)}{;}{a}{}{,}
\usepackage{txfonts}
\usepackage[breaklinks,pdftex]{hyperref}
\begin{document} 

   \title{\textit{Spitzer} mid-infrared point sources in the fields 
     of\\ nearby galaxies
     \thanks{Based on observations made with the \textit{Spitzer} Space
       Telescope, which is operated by the Jet Propulsion Laboratory,
       California Institute of Technology under a contract with NASA.}
     \thanks{Tables 2$-$6 are only available in electronic form at
       the CDS via anonymous ftp to cdsarc.u-strasbg.fr (130.79.128.5)
       or via \url{http://cdsweb.u-strasbg.fr/cgi-bin/qcat?J/A+A/}}.}

   \author{S. J. Williams
          \inst{1,2}
          \and
	  A. Z. Bonanos
	  \inst{2}}

   \institute{Department of Physics, University of Crete, GR-71003, Heraklion, 
             Greece
             \email{williams@physics.uoc.gr}
         \and
             IAASARS, National Observatory of Athens, GR-15236 Penteli, Greece}

   \date{Received ; accepted }

\authorrunning{Williams et al.}
\titlerunning{Mid-infrared Point Sources}

\abstract
{}
{To complement the study of transient phenomena and to assist 
  subsequent observations in the mid-infrared, we extract point source
  photometry from archival mosaics of nearby
  galaxies with high star formation rates within 4 Mpc.}
{Point spread function photometry was performed on sources detected in both 
  \textit{Spitzer} IRAC 3.6 $\mu$m and 4.5 $\mu$m bands at greater than
  3$\sigma$ above background. These data were
  then supplemented by aperture photometry in the IRAC 5.8 $\mu$m and
  8.0 $\mu$m bands conducted at the positions of the shorter wavelength 
  sources. For sources with no detected object in the longer
  wavelengths, we estimated magnitude limits based on the local 
  sky background.}
{We present \textit{Spitzer} IRAC mid-infrared point source catalogs
  for mosaics covering the fields of the nearby ($\lesssim$4 Mpc) galaxies 
  NGC 55, NGC 253, NGC 2366, NGC 4214, and NGC 5253. We detect a total of
  20159 sources in these five fields. The individual galaxy point
  source breakdown is the following: NGC 55, 8746 sources; NGC 253, 9001
  sources; NGC 2366, 505 sources; NGC 4214, 1185 sources; NGC 5253, 722
  sources. The completeness limits of the full catalog vary with bandpass 
  and were found to be $m_{3.6}=18.0$, $m_{4.5}=17.5$, $m_{5.8}=17.0$, and
  $m_{8.0}=16.5$ mag. For all galaxies, this corresponds to detection of
  point sources brighter than $M_{3.6}=-10$.
  These catalogs can be used as a reference for stellar
  population investigations, individual stellar object studies, and in
  planning future mid-infrared observations with the James Webb Space 
  Telescope.}
{}


\maketitle

%

\section{Introduction}

Understanding the resolved stellar populations of nearby galaxies 
is vital to understanding the integrated light of
more distant galaxies. Extrapolating the properties of local, resolved
stellar populations is one method by which the properties of
the first generation of stars can be investigated \citep{sob15}. 
Other examples of the benefits of studying 
resolved stellar populations 
range from constraining the formation and evolution of compact
objects \citep{leh14} to identifying the progenitors and
environments of supernovae (SNe) and other transients 
\citep[e.g.,][]{pri08}.

The examination of individual stars in nearby galaxies can also 
provide insight into stellar evolutionary theory. 
The most luminous point sources will typically be the most 
massive evolved stars \citep{hos14}. The characterization of 
these sources sheds light on an otherwise poorly understood area of
modern stellar astrophysics. The later stages of evolution for
massive stars can produce copious amounts of intervening 
circumstellar material, making observations difficult or
impossible in optical bands \citep{gva10,wac10,kha15a}. 
Taking advantage of the largely
untapped potential for individual stellar investigations in
the local universe with mid-infrared (IR) \textit{Spitzer} 
archival data, several forays into the stellar content of the
Local Group members \citep{mou08,bon09,tho09,bon10} were 
performed. Investigations subsequently expanded to more distant
galaxies. For instance, \citet{kha13} searched for $\eta$ Carinae analogs,
uncovered an emerging class of evolved massive stars 
\citep{kha15b}, and later discovered several $\eta$ Carinae
analogs \citep{kha15a}. \citet{wil15} extracted photometry from
archival \textit{Spitzer} Infrared Array Camera (IRAC)
images of the galaxy M83 and used color-magnitude diagrams 
(CMD) of 3.6 $\mu$m and 4.5 $\mu$m photometry to identify
red supergiant (RSG) candidates. Mid-IR photometric candidates  
of this kind have been shown via spectroscopic follow up 
to indeed be RSGs (late-type stars with zero age main 
sequence mass between $\sim8$ and $\sim30$ $M_{\odot}$) by 
\citet{bri14,bri15}.

The aim of our work here is to continue to stockpile 
mid-IR point source catalogs for nearby galaxies and their fields.
\citet{kha15c} presented catalogs of point sources in seven
nearby galaxies (M33, M81, NGC 247, NGC 300, NGC 2403, NGC 6822,
and NGC 7793) from mid-IR \textit{Spitzer} archival observations.
These catalogs can be used for stellar population investigations,
individual stellar object studies, and as a reference for planning 
future mid-IR observations with the James Webb Space Telescope 
\citep[JWST;][]{gar06}. 

We start with a discussion concerning
our target selection in Section \ref{targets}, 
then we outline our photometric analysis
procedure in Section \ref{photometry}, and describe the 
catalogs in Section \ref{catalogs}. Finally, we wrap up the 
paper in Section \ref{summary} with a summary of possible 
applications of sources found in this work.


\section{Target selection}\label{targets}

Galaxies were chosen that met the following criteria: 1) galaxies
for which point source catalogs from \textit{Spitzer} are not yet 
published, 2)
galaxies close enough ($\lesssim$4 Mpc) that the resolution 
allows measurements of individual stars, 3) galaxies outside the Local 
Group, and 4) galaxies with a star formation rate listed in 
\citet{ken08} with log ($L_{H\alpha (ergs/s)}$)>40. In order to ensure 
the homogeneity of our data set, we also limited our sample to 
galaxies within 4 Mpc that exist in the Local Volume 
Legacy survey \citep[LVL;][]{dal09}, which has a
pixel scale of 0$\farcs$75.
There are a total of five galaxies meeting the above selection criteria: 
NGC 55, NGC 253, NGC 2366, NGC 4214, and NGC 5253. 
The properties of these galaxies are summarized in Table 
\ref{table:gal_props} and their characteristics are described below. 

NGC 55 is an edge-on
spiral galaxy associated with the Sculptor Group that likely is
part of a gravitationally bound pair with the galaxy NGC 300 
\citep{van06} at a Tip of the Red Giant Branch (TRGB) 
distance of 2.1$\pm$0.1 Mpc
\citep{tan11}. The proximity and size of NGC 55 make it a
prime target for studies of individual stars. For example,
\citet{cas12} spectroscopically studied several young,
massive B-type supergiants likely evolving toward the RSG
phase in NGC 55. 

At the core of the Sculptor Group lies NGC 253,
shown in the IR to be a barred spiral of classification SAB(s)c
\citep{jar03} at 3.5$\pm$0.1 Mpc \citep{rad11}. Resolved individual
stars in NGC 253 have been used to study the disk structure and
halo \citep{gre14,bai11}, but evolved massive stars remain 
largely unstudied.

NGC 2366 (DDO 42) is the closest member of the M81 Group of galaxies
at a distance of $\sim3.2$ Mpc \citep{kar02}. It is an irregular 
galaxy that is a companion to NGC 2403, one of the main members 
of the M81 Group. One exceptional massive evolved star was revealed 
to be a luminous blue variable in outburst by \citet{dri01},
while the broader stellar population of 
\textit{Hubble Space Telescope} (\textit{HST}) 
imaging fields revealed a few evolved massive star candidates
\citep{thu05,tik08}. 

NGC 4214 is a dwarf irregular Magellanic-type 
galaxy in the direction of the Canes Venatici cloud at a distance 
of 2.7$\pm$0.3 Mpc \citep{dro02}. \textit{HST} observations of NGC 
4214 \citep{ube07a} resolved individual massive evolved stars 
and found a steep initial mass function for stars above 
$20~M_{\odot}$ with several RSG and blue supergiant candidates 
\citep{ube07b}. 

Finally, NGC 5253 is a blue 
compact dwarf galaxy in the M83 Group at a distance of 3.6$\pm$0.2
Mpc \citep{sak04}. The central region of NGC 5253 shows evidence 
for a recent starburst and large numbers of young massive stars
\citep{cro99}.


\section{Photometry}\label{photometry}

The photometric analysis of the data made use of the DAOPHOT 
\citep{ste87,ste92} package in IRAF\footnote{IRAF is distributed
by the National Optical Astronomy Observatory, which is operated by 
the Association of Universities for Research in Astronomy (AURA) Inc.,
under cooperative agreement with the National Science Foundation.}. 
We began by following the routine described in \citet{wil15} for 
the 3.6 $\mu$m and 4.5 $\mu$m mosaics separately. We first detected
point sources, then constructed a point-spread function (PSF) from
bright, isolated stars. In the LVL catalog, the 3.6 $\mu$m and 
4.5 $\mu$m mosaics for the chosen galaxies do not overlap exactly,
meaning it was necessary to use different PSF stars for the two 
bandpasses. PSF photometry was then performed on the entirety of the
point source catalogs. For the PSF stars, aperture photometry was
converted to the Vega system using the aperture corrections 
and zero-point fluxes listed in the IRAC Instrument 
Handbook\footnote{\url{http://irsa.ipac.caltech.edu/data/SPITZER/docs/irac/iracinstrumenthandbook/}}. 
For the aperture photometry, we used a 3$\farcs$6 aperture radius,
with the sky annulus beginning at the edge of the
aperture radius and ending at a radius of 8$\farcs$4.
The offset between this calibrated aperture photometry 
and the PSF photometry was then applied to the entire data set.
The Vega-calibrated PSF data sets in the 3.6 $\mu$m and 4.5 $\mu$m
catalogs were then coordinate-matched using J. D. Smith's 
{\tt match\_2d.pro}\footnote{\url{http://tir.astro.utoledo.edu/jdsmith/code/idl.php}}.
\citet{kha15c} investigated the best coordinate matching radius
for their \textit{Spitzer} 3.6 $\mu$m and 4.5 $\mu$m catalogs.
They found that the majority of sources ($>$90\%) are matched within one half
pixel, with virtually all reasonable matches lying within one pixel.
Therefore, the matching criteria employed here uses a radius of 
one pixel, or 0$\farcs$75. The final coordinates adopted for our 
catalog come from the image header World Coordinate Systems in
3.6 $\mu$m mosaics.

After coordinate matching the two shortest wavelength bands, we
proceed to add to the catalog aperture photometry in
the 5.8 $\mu$m and 8.0 $\mu$m at the coordinates determined from
the 3.6 $\mu$m band. For instances where there is no object
in either of these bands, we use the local sky background,
which was measured at the position of the detection
in the 3.6 $\mu$m band, to
determine the limit of the magnitude. These instances
amounted to 3374 magnitude limits in the 5.8 $\mu$m band
and 5107 similarly measured objects in the 8.0 $\mu$m band.
These data were also calibrated by applying appropriate constants 
from the IRAC Instrument Handbook, with the same extraction 
apertures and sky annuli as for PSF stars in order 
to calibrate the data for the 3.6 $\mu$m and 4.5 $\mu$m bands 
(an aperture radius of 3$\farcs$6 and sky annulus of 4$\farcs$8).

The LVL survey \citep{dal09} created mosaic images of galaxies
constructed from a number of exposures in the fields of each
galaxy. Because of this, exposure times differed on a pixel-by-pixel
basis. Their ``coverage maps'' for each galaxy list a representative
exposure time for each pixel in the mosaic. For NGC 55, NGC 2366,
and NGC 4214, these exposure times are on average 500 seconds for
pixels close to the particular galaxy. For NGC 253, the exposure 
times range between 300 and 500 seconds, with most pixels being
closer to 500 seconds. However, for the NGC 5253 mosaic, the 
average exposure times are closer to 300 seconds. This may, in some
regard, mean the catalog for NGC 5253 is not as deep as those
for the other galaxies.

\section{Point Source Lists}\label{catalogs}

The total number of 
point sources is 20159, with 8746 sources in NGC 55 (Table 
\ref{table:NGC55}), 9001 sources in NGC 253 (Table \ref{table:NGC253}), 
505 sources in NGC 2366 (Table \ref{table:NGC2366}), 
1185 sources in NGC 4214 (Table \ref{table:NGC4214}), and
722 sources in NGC 5253 (Table \ref{table:NGC5253}).
Tables \ref{table:NGC55} through \ref{table:NGC5253} are available
in their entirety in the online version of the journal, but
a portion of the point source catalog for each individual galaxy
is shown here. The columns of the tables are uniform and include the
RA and Dec (J2000) in degrees followed by the PSF magnitudes and
associated uncertainties for the 3.6 $\mu$m and 4.5 $\mu$m bands,
with the last four columns containing the aperture photometry and
uncertainties in the 5.8 $\mu$m and 8.0 $\mu$m bandpasses. 
Our uncertainty estimates come from the DAOPHOT PSF-fitting
or sky annuli for aperture photometry and do not include 
calibration uncertainties. According to the 
\textit{Spitzer} instrument handbook, typical calibration 
uncertainties are on the order of $\sim$0.03 mag.
Each table is sorted by brightness in the 3.6 $\mu$m band. For any
measurement in the 5.8 $\mu$m or 8.0 $\mu$m bands that represents
a non-detection, an uncertainty of ``9.99'' is listed in order to
denote the estimate as a limit. 

\begin{figure*}
\centering
\includegraphics[width=17cm]{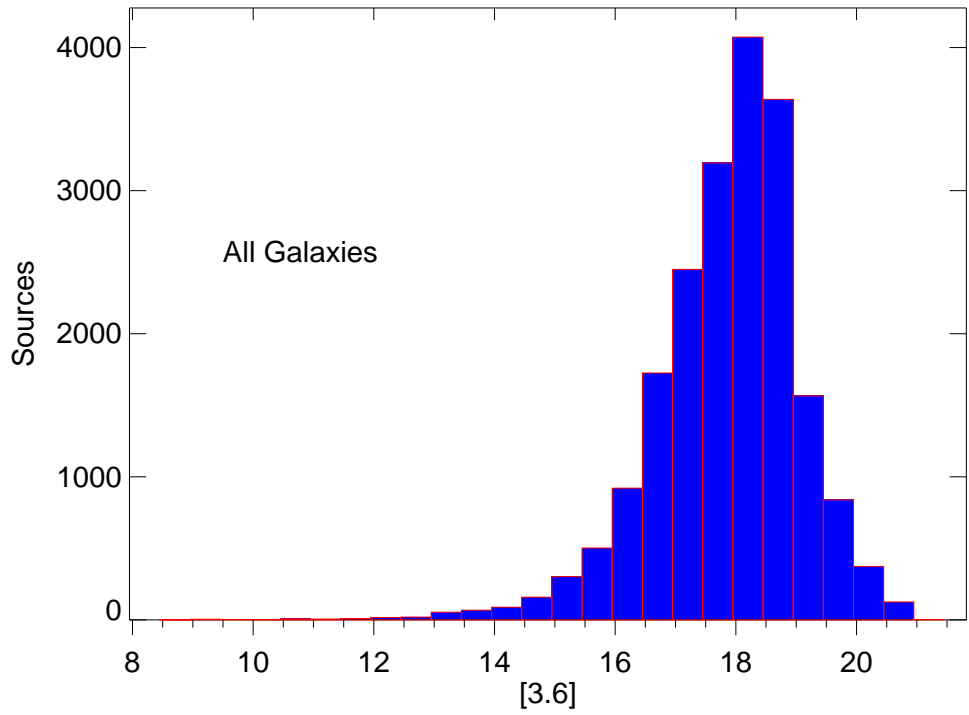}
\caption{Histogram showing the detected sources per half magnitude bin
  for the entire sample of 20159 sources across all five galaxies in the 
  \textit{Spitzer} IRAC 3.6 $\mu$m band. The limiting magnitude is estimated
  to be where the histogram turns over, at $m_{[3.6]}$=18.0.}
\label{Band1_hist}
\end{figure*}

\begin{figure*}
\centering
\includegraphics[width=17cm]{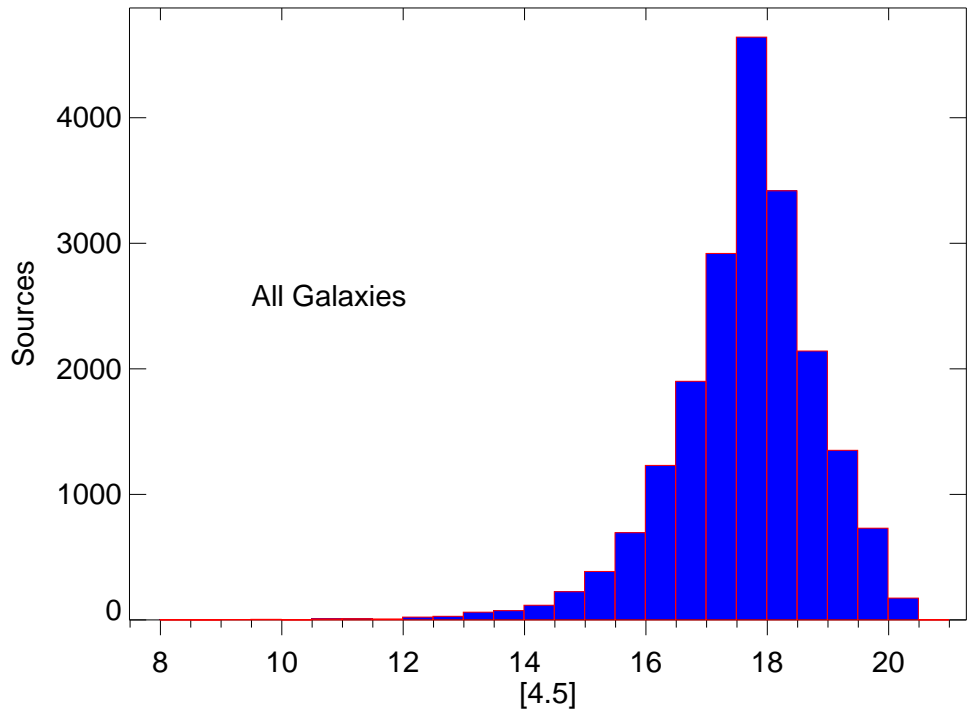}
\caption{Same as Figure \ref{Band1_hist} but in the 
  \textit{Spitzer} IRAC 4.5 $\mu$m band. The limiting magnitude is estimated
  to be where the histogram turns over, at $m_{[4.5]}$=17.5.}
\label{Band2_hist}
\end{figure*}

\begin{figure*}
\centering
\includegraphics[width=17cm]{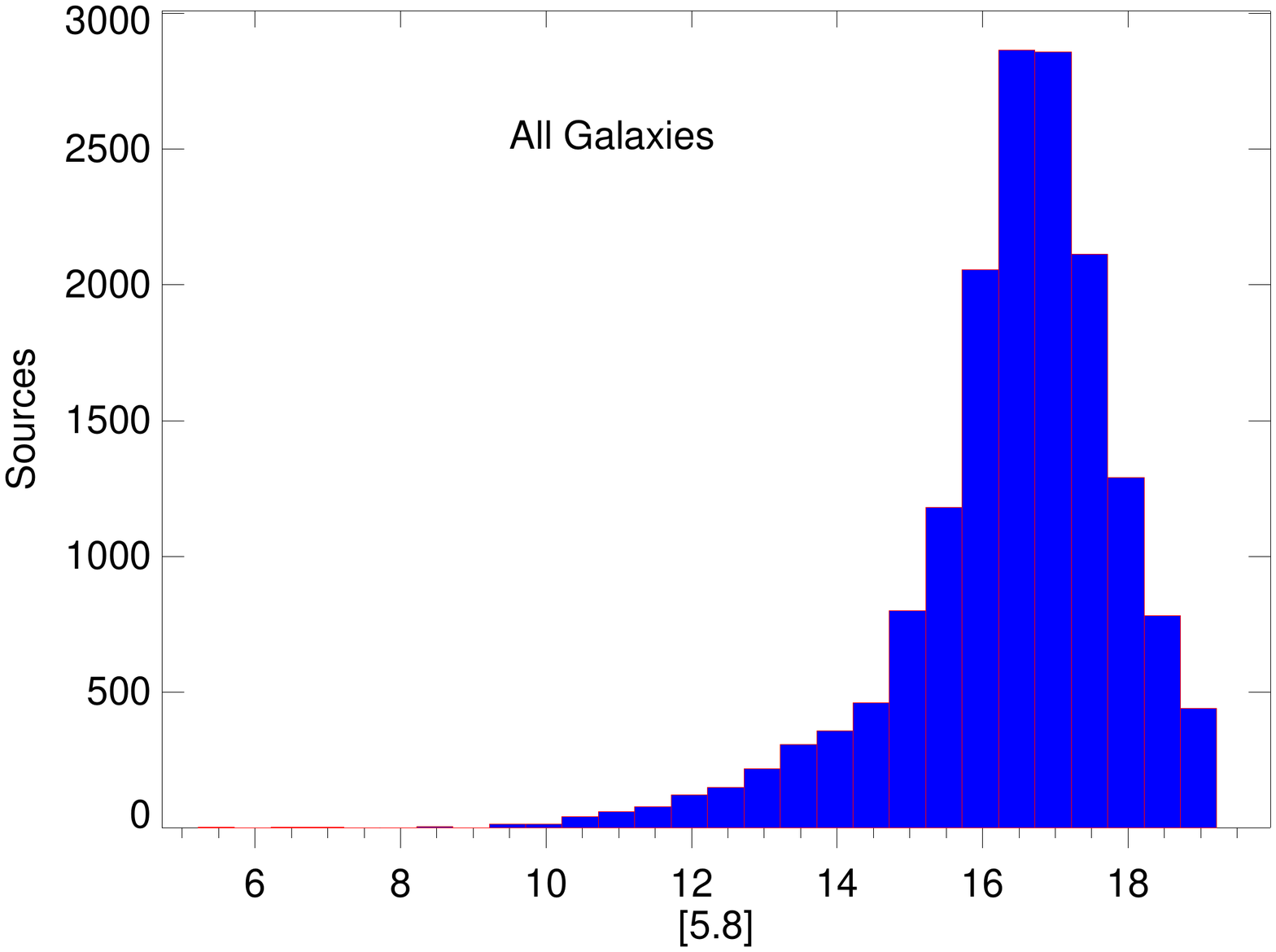}
\caption{Same as Figure \ref{Band1_hist} but in the 
  \textit{Spitzer} IRAC 5.8 $\mu$m band, with 3949 sources removed that
  were not recovered in this bandpass. The limiting magnitude is estimated
  to be where the histogram turns over, at $m_{[5.8]}$=17.0.}
\label{Band3_hist}
\end{figure*}

\begin{figure*}
\centering
\includegraphics[width=17cm]{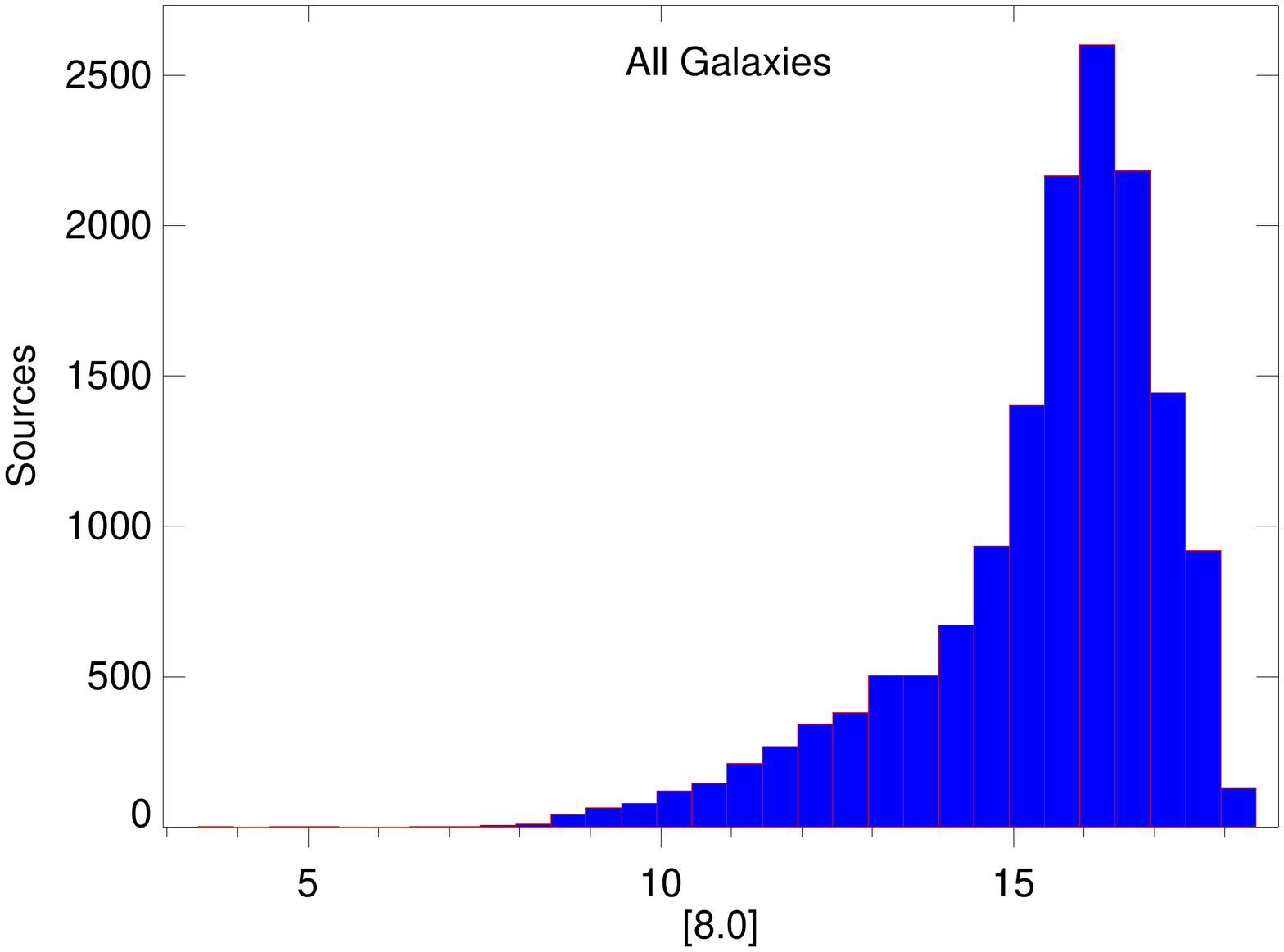}
\caption{Same as Figure \ref{Band1_hist} but in the 
  \textit{Spitzer} IRAC 8.0 $\mu$m band, with 5022 sources removed that
  were not recovered in this bandpass. The limiting magnitude is estimated
  to be where the histogram turns over, at $m_{[8.0]}$=16.5.}
\label{Band4_hist}
\end{figure*}

The catalogs list any point source meeting our criteria from 
section \ref{photometry}. To estimate the completeness of
the catalogs, we constructed a histogram of photometry from 
all galaxies in each bandpass for half-magnitude bins. 
Because the number of sources should scale with
the volume sampled, and the volume sampled should scale according
to the limiting luminosity,
one expects a histogram to roughly follow a power law of order 3. 
Therefore, 100\% completeness is lost where the histogram drops 
off from this power law. Figures \ref{Band1_hist} through 
\ref{Band4_hist} show the histogram distributions for the
3.6 $\mu$m through 8.0 $\mu$m bandpasses. In each case, the
drop off in the increase of detected sources is obvious, and
we estimate the limiting magnitudes of the catalogs to be
$m_{[3.6]}=18.0$, $m_{[4.5]}=17.5$, $m_{[5.8]}=17.0$, and $m_{[8.0]}=16.5$, 
corresponding to an average absolute magnitude for
the galaxies of $M_{3.6}=-9.3$, $M_{4.5}=-9.8$, $M_{5.8}=-10.3$, and
$M_{8.0}=-10.8$.
These limits are consistent with those found for other
nearby galaxies \citep{kha15c}.
 
We can compare the number of sources
detected for our galaxies to the numbers detected by \citet{kha15c}.
Initially, it is clear that there are fewer sources
detected in our galaxies than in those from the 
study by \citet{kha15c}. This is likely due to the interplay
of several factors. First, the mosaics analyzed by \citet{kha15c}
typically cover a larger area of the sky than those studied here 
(compare Table 1 in \citealt{kha15c} to our 
Table \ref{table:gal_props}). Second, a greater distance 
to a particular galaxy will mean fewer detected point sources, 
as is the case with NGC 7793 (at a distance of $\sim4.1$ Mpc) 
in \citet{kha15c}. The last main factor in the difference is
due to the orientation and physical size of the galaxy in question.
Both NGC 55 and NGC 253, the galaxies with the two highest
number of point sources in our study, are highly inclined to
our line of sight. This means less projected area on the sky
for these galaxies, and fewer detected point sources. These
galaxies are also intrinsically smaller in size, containing
fewer stars, than M81, for example. In short, intrinsically
smaller galaxies, at higher inclination angles to our 
line of sight, at on average larger distances, with mosaics
covering smaller areas on the sky mean fewer detected
point sources.

\begin{figure*}
\centering
\includegraphics[width=0.9\paperwidth]{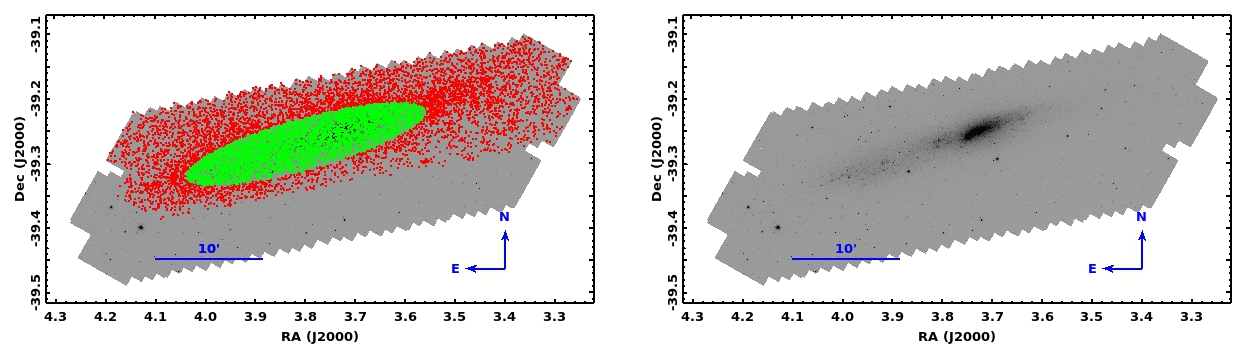}
\caption{\textit{Spitzer} IRAC 3.6 $\mu$m image of NGC 55 both with 
  and without . Axes are
  labeled in units of degrees. Red points show
  locations of sources, while green dots show sources within our estimate
  of the galaxy's isophotal borders based initially upon the Kron radius 
  described in the 2MASS Large Galaxy Atlas \citep{jar03}. 
  For reference, also shown are a compass indicating the orientation 
  of the image and a bar of length 10 arc minutes. }
\label{NGC55_images}
\end{figure*}

\begin{figure*}
\centering
\includegraphics[width=17cm]{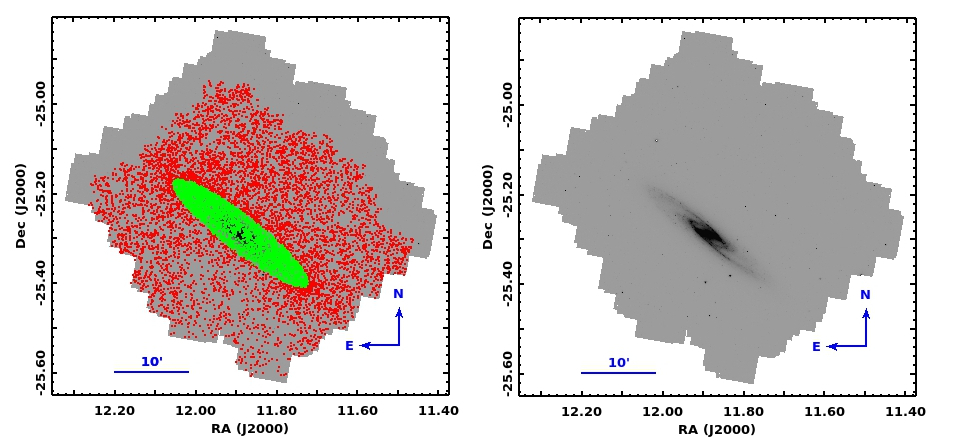}
\caption{\textit{Spitzer} IRAC 3.6 $\mu$m image of NGC 253. Axes are
  labeled in units of degrees. Red points show
  locations of sources, while green dots show sources within our estimate
  of the galaxy's isophotal borders based initially upon the Kron radius 
  described in the 2MASS Large Galaxy Atlas \citep{jar03}.
  For reference, also shown are a compass indicating
  the orientation of the image and a bar of length 10 arc minutes.}
\label{NGC253_images}
\end{figure*}

\begin{figure*}
\centering
\includegraphics[width=17cm]{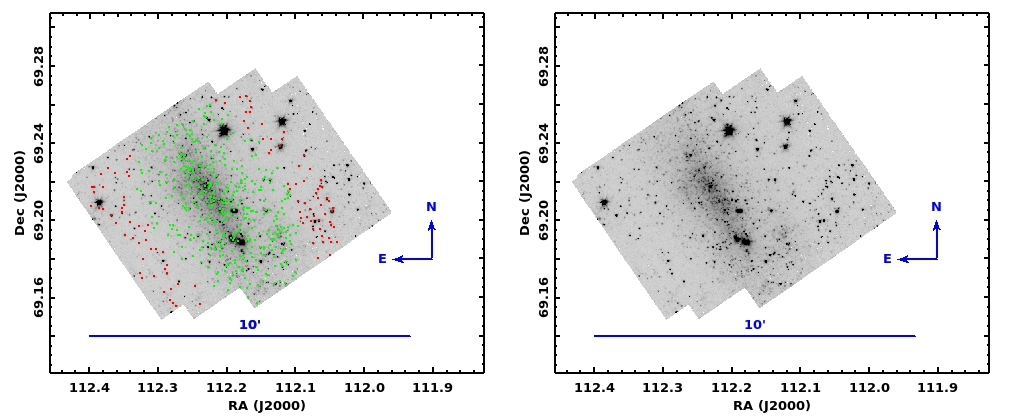}
\caption{\textit{Spitzer} IRAC 3.6 $\mu$m image of NGC 2366. Axes are
  labeled in units of degrees. Red points show
  locations of sources, while green dots show sources within our estimate
  of the galaxy's isophotal borders based initially on isophotes
  described in \citet{dev91}. For reference, also shown are a compass indicating
  the orientation of the image and a bar of length 10 arc minutes.}
\label{NGC2366_images}
\end{figure*}

\begin{figure*}
\centering
\includegraphics[width=17cm]{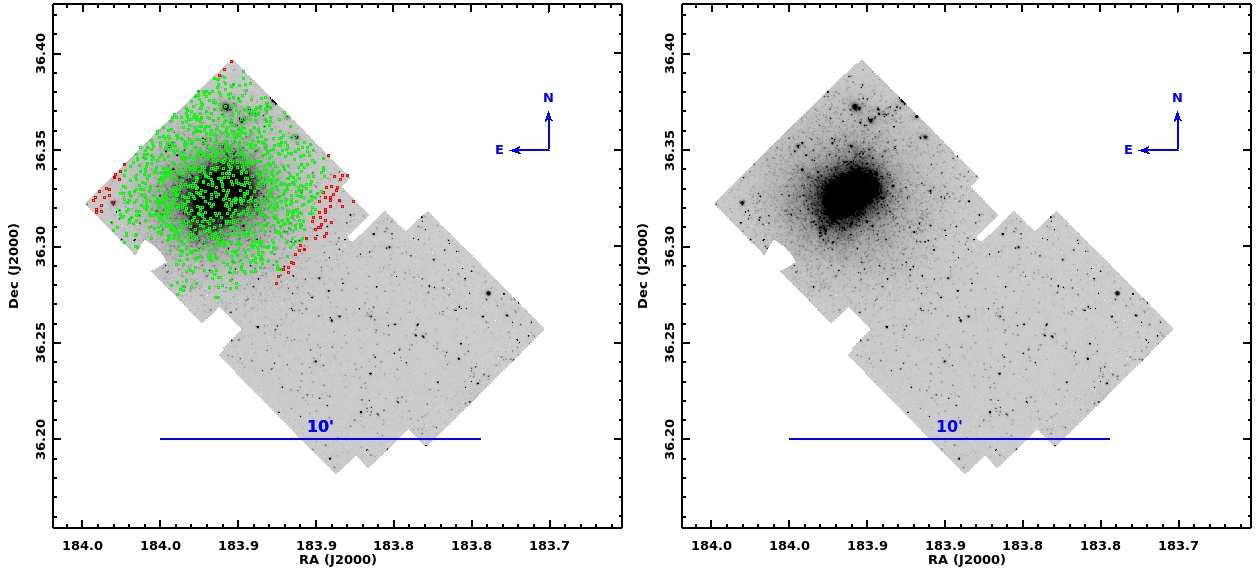}
\caption{\textit{Spitzer} IRAC 3.6 $\mu$m image of NGC 4214. Axes are
  labeled in units of degrees. Red points show
  locations of sources, while green dots show sources within our estimate
  of the galaxy's isophotal borders based initially upon the Kron radius 
  described in the 2MASS Large Galaxy Atlas \citep{jar03}.
  For reference, also shown are a compass indicating
  the orientation of the image and a bar of length 10 arc minutes.
  It should be noted that the 4.5 $\mu$m image only overlaps with the
  galaxy itself, thus the large portion of this image with no labeled 
  sources.}
\label{NGC4214_images}
\end{figure*}

\begin{figure*}
\centering
\includegraphics[width=17cm]{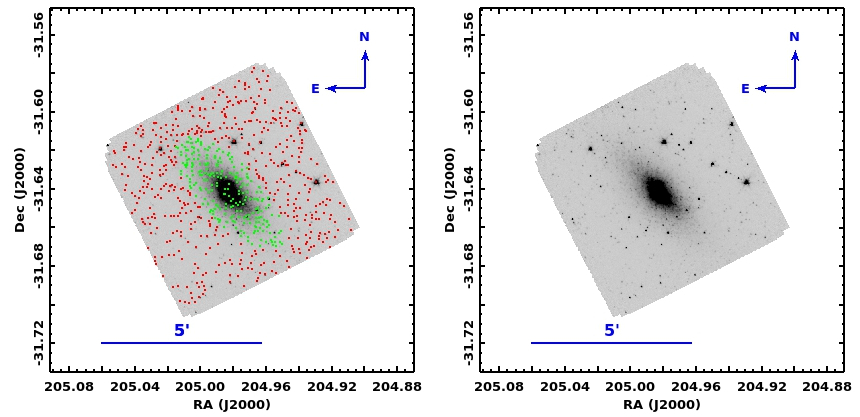}\label{NGC5253_points}
\caption{\textit{Spitzer} IRAC 3.6 $\mu$m image of NGC 5253. Axes are
  labeled in units of degrees. Red points show
  locations of sources, while green dots show sources within our estimate
  of the galaxy's isophotal borders based initially upon the Kron radius 
  described in the 2MASS Large Galaxy Atlas \citep{jar03}.
  For reference, also shown are a compass indicating
  the orientation of the image and a bar of length 5 arc minutes.}
\label{NGC5253_images}
\end{figure*}

\citet{wil15} investigated the stellar content of M83, removing
sources that were not associated with the galaxy or were known
to not be stellar in nature (proper motion stars, clusters, 
supernova remnants, and so on). Here, we are interested in
simply cataloging the point sources matched between the two
shortest wavelength IRAC bands. To illustrate the locations of
detected sources in our catalog, we present in Figures 
\ref{NGC55_images} through \ref{NGC5253_images} the 3.6 $\mu$m 
image of each galaxy with the positions of detected sources 
overplotted, followed by an image of the galaxy without sources 
shown. For each galaxy, red points denote the location of point
sources not likely to be associated with the galaxy, while
green points demarcate the regions inside isophotes for the
Kron radius for all galaxies but NGC 2366, where the isophote
described in \citet{dev91} was used.
Consequently, the green points are more probably
actual sources associated with each galaxy. 

\begin{figure*}
\centering
\includegraphics[width=17cm]{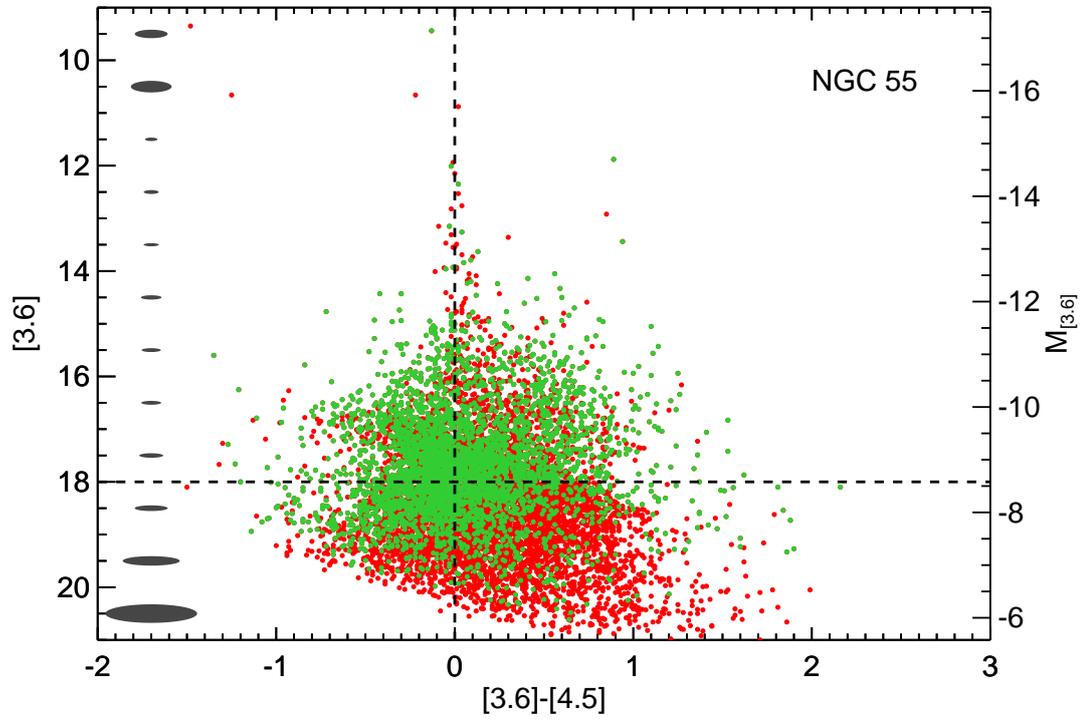}
\caption{Color-magnitude diagram for sources in the field of NGC 55. All
  photometrically measured sources are marked with a red dot, while those
  residing within the isophotal limits of the galaxy are represented by
  green dots. Overplotted are lines denoting the limiting magnitude of 18.0
  and a [3.6]-[4.5] color of 0. On the left are ellipses corresponding 
  to the average uncertainty for all points in every one magnitude bin.}
\label{NGC55_cmd}
\end{figure*}

\begin{figure*}
\centering
\includegraphics[width=17cm]{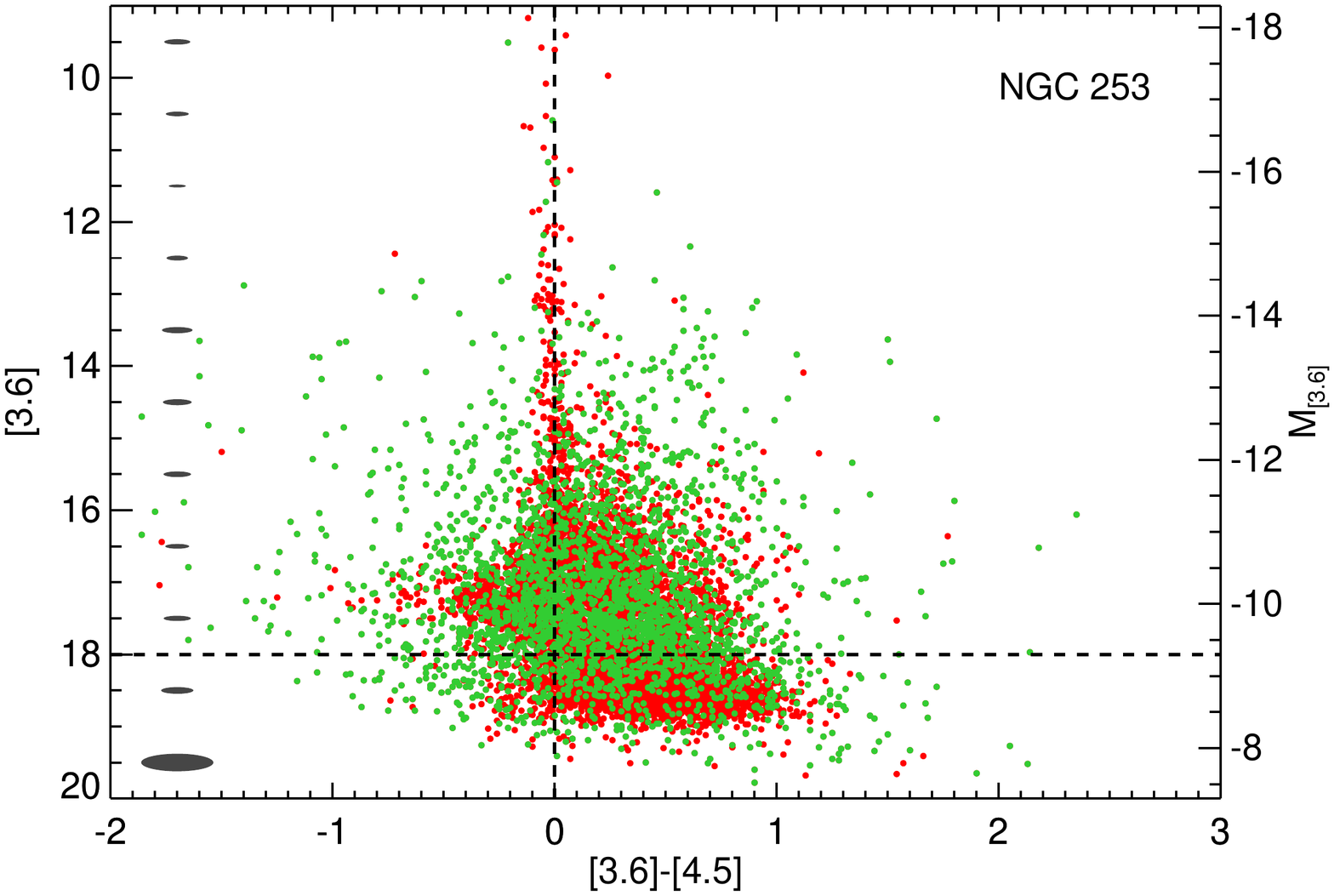}
\caption{Same as Figure \ref{NGC55_cmd} but for sources in the 
field of NGC 253.}
\label{NGC253_cmd}
\end{figure*}

\begin{figure*}
\centering
\includegraphics[width=17cm]{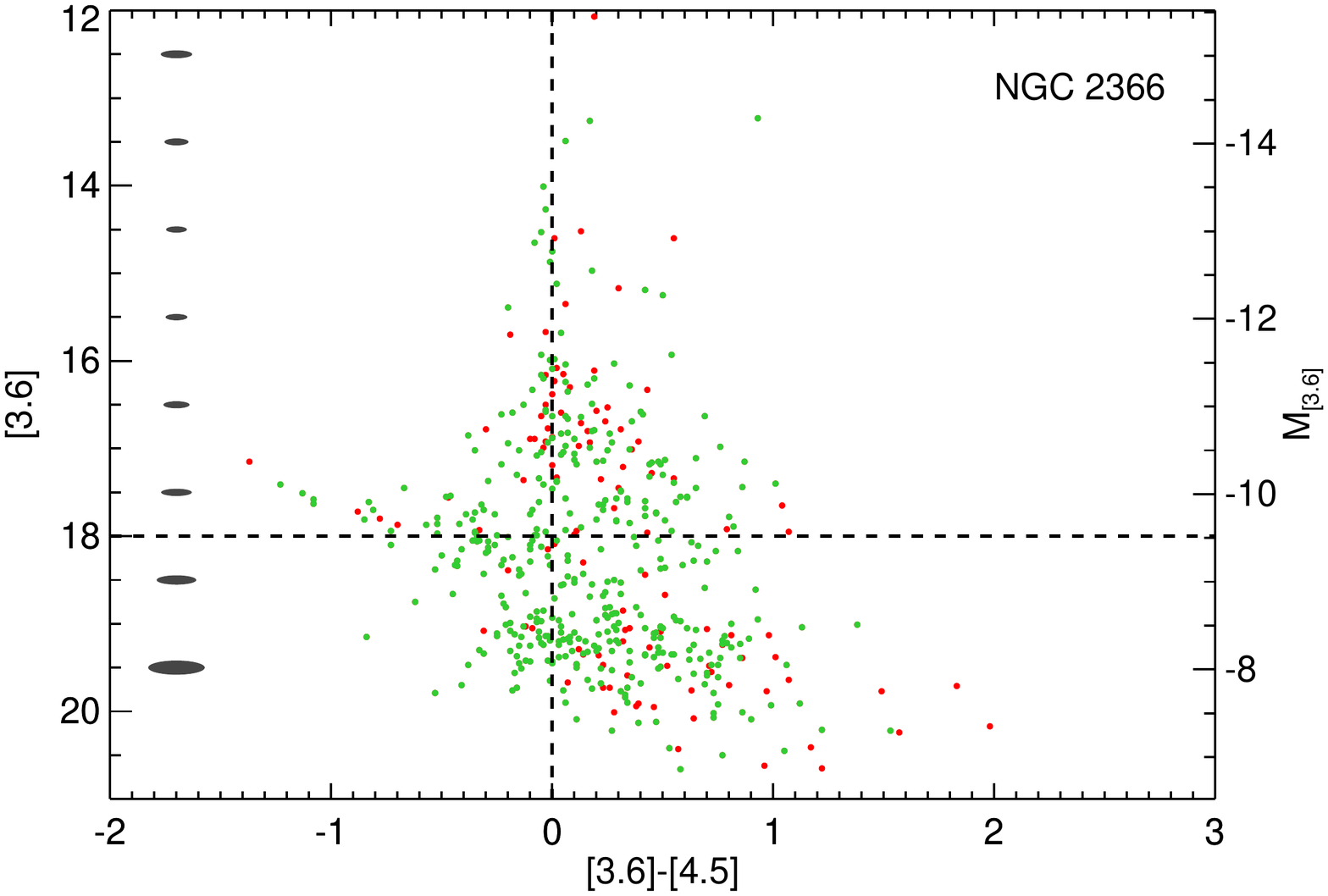}
\caption{Same as Figure \ref{NGC55_cmd} but for sources in the 
field of NGC 2366.}
\label{NGC2366_cmd}
\end{figure*}

\begin{figure*}
\centering
\includegraphics[width=17cm]{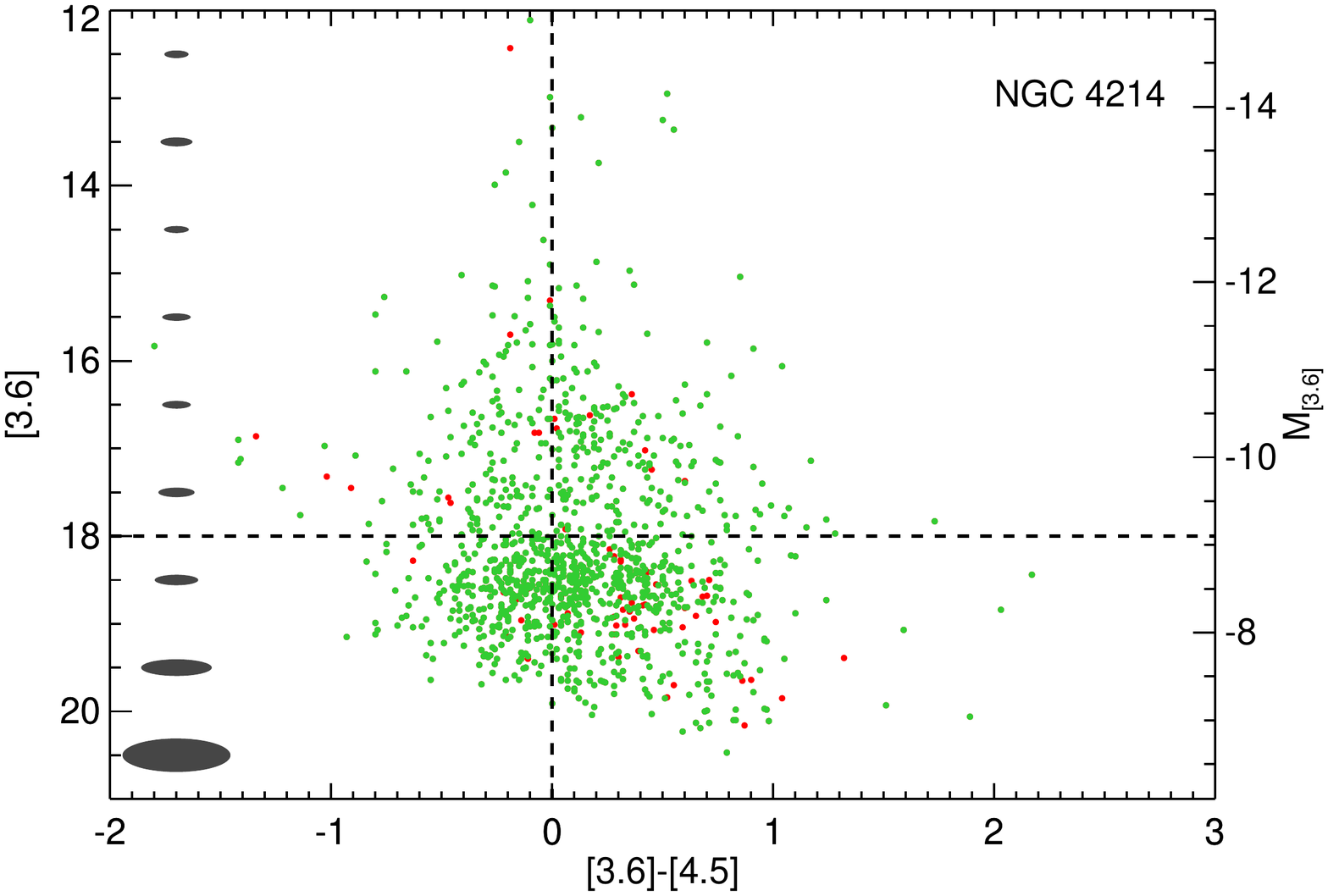}
\caption{Same as Figure \ref{NGC55_cmd} but for sources in the 
field of NGC 4214.}
\label{NGC4214_cmd}
\end{figure*}

\begin{figure*}
\centering
\includegraphics[width=17cm]{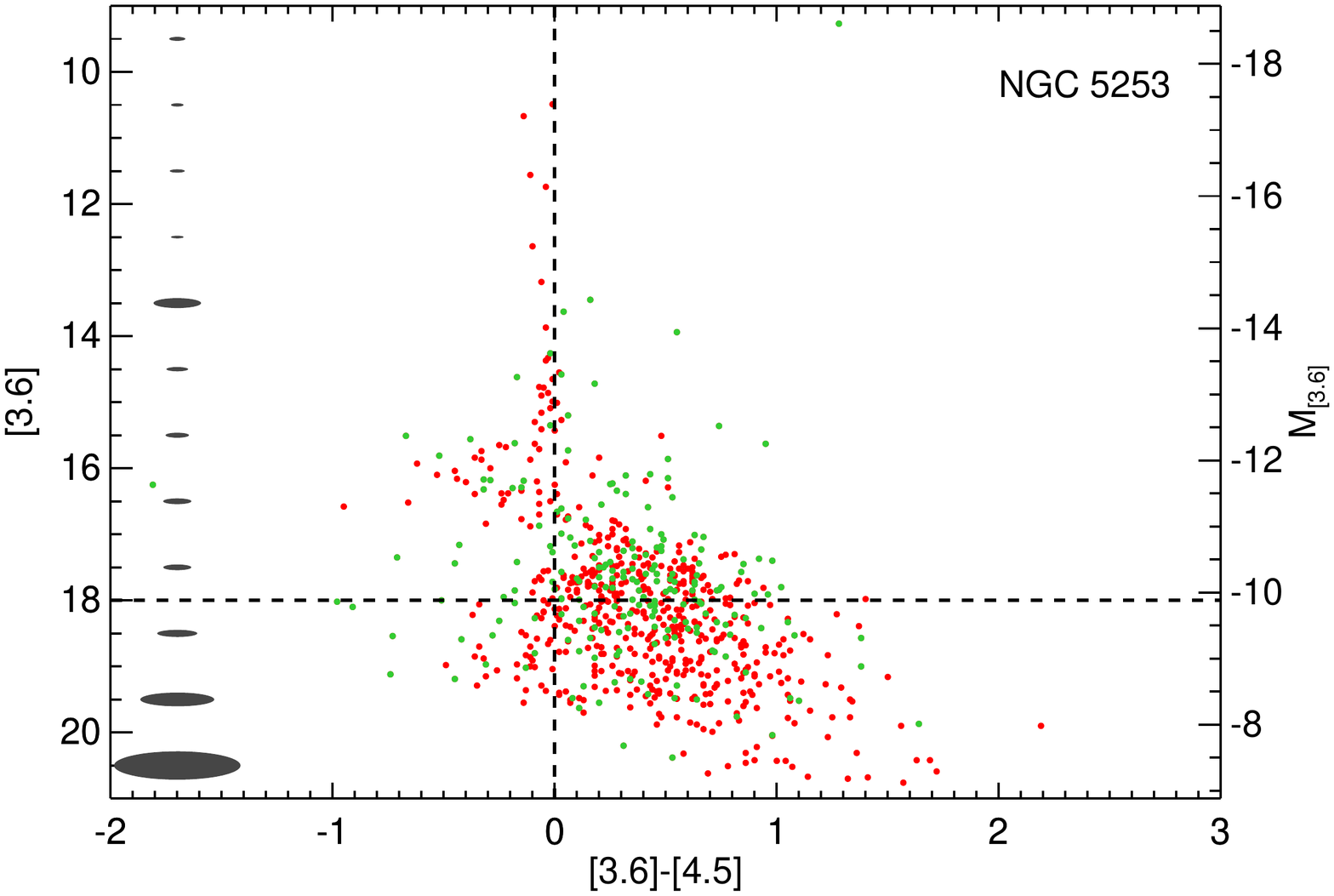}
\caption{Same as Figure \ref{NGC55_cmd} but for sources in the 
field of NGC 5253.}
\label{NGC5253_cmd}
\end{figure*}

Figures \ref{NGC55_cmd} through \ref{NGC5253_cmd} show
color-magnitude diagrams (CMDs) of [3.6] versus [3.6]$-$[4.5] 
for each galaxy. The points follow the color scheme from the
images showing each galaxy, with green points representing
sources within the isophotes for with each galaxy, and red
points sources in the field of each galaxy. For each 
magnitude bin, the average uncertainty was computed for both
the [3.6] magnitude and [3.6]$-$[4.5] color, and is
plotted as an ellipse on the left hand side of the plot.
For reference, lines are drawn to show the completeness
limit of $m_{[3.6]}=18.0$ mag and the color [3.6]$-$[4.5] = 0,
where most stars lie in the plot. In terms of absolute magnitudes,
the [3.6] limit corresponds to detections of all
objects brighter than $M_{3.6}=-10$ in every galaxy.
One other interesting
feature in the plot was explained by \citet{kha15c}. For
both NGC 55 and NGC 253, the two galaxies with the most
sources of those studied in this work, one can see a
``break'' at about the same magnitude as the completeness
limit. This feature is representative of the two different
populations in the photometry, one physically associated
with the galaxy, and another simply sharing the same region
of sky. \citet{kha15c} calls the second population
``contaminants''. This population contains foreground stars
near the [3.6]$-$[4.5] color of 0, some Milky Way Halo 
late-type giants and 
unresolved background galaxies in other regions
of the CMD. Red supergiants (RSGs) have
been empirically shown to exist in the region of negative 
color space \citep{bon09,bon10}. 
This is due to a suppression of the flux
in the 4.5 $\mu$m photometry owing to molecular bands,
most prominently CO \citep{ver09}. However, spectroscopic
investigations into the individual stellar components of
Local Group dwarf irregular galaxies have discovered 
$\sim15$\% of new RSGs at positive colors in CMDs for
those galaxies \citep{bri14,bri15}.

As an example of the utility of our catalog, the analysis 
of NGC 253 includes an RSG recently shown to be
associated with the ultraluminous X-ray (ULX) source, 
RX J004722.4-252051 \citep{hei15}. Our photometry reveals
a color of $[3.6]-[4.5] = -0.43\pm0.36$, placing this object
squarely in the empirical region where RSGs are known to
exist. The absolute magnitude of this object is $M_{3.6}=-9.85$
again lying in the empirical RSG region, adding further, 
independent support to the 
authors' conclusions that the counterpart is a red supergiant.

\section{Summary}\label{summary}

We present \textit{Spitzer} point source catalogs 
for 20159 total sources with photometric 
measurements in the four IRAC bandpasses: 3.6 $\mu$m,
4.5 $\mu$m, 5.8 $\mu$m, and 8.0 $\mu$m. The point sources
have been extracted from the LVL \citep{dal09} mosaics
in the fields of the galaxies NGC 55 (8746 sources), 
NGC 253 (9001 sources), NGC 2366 (505 sources),
NGC 4214 (1185 sources), and NGC 5253 (722 sources).
For all galaxies, this corresponds to detection of
point sources brighter than $M_{3.6}=-10$.

The catalogs presented here can be used for a number of
future works. Our particular interest lies in identifying 
dusty evolved massive stars such as RSGs, 
supergiant B[e] stars, luminous blue variables, and 
rare Fe emission stars for follow-up 
spectroscopy. Such identifications will enable a more
detailed study of the stellar populations of these
galaxies. This archive can also be used to plan
future mid-IR observations conducted with JWST to identify
and understand the nature of supernova progenitors
and supernova remnants \citep{leo10}. Finally, it will prove 
useful for multiwavelength studies of targets in these fields, 
such as X-ray binaries, as demonstrated by our identification
of a mid-IR counterpart for \textbf{t}he ULX in NGC 253.

\begin{acknowledgements}

We wish to thank the anonymous referee for a careful reading
of the text and suggesting improvements to clarify the
presentation of the material.
SJW and AZB acknowledge funding by the European Union
(European Social Fund) and National Resources under the
``ARISTEIA'' action of the Operational Programme ``Education
and Lifelong Learning'' in Greece. 
This research has made use of the VizieR catalogue access tool, 
CDS, Strasbourg, France. This work is based [in part] on observations 
made with the
\textit{Spitzer Space Telescope}, which is operated by the Jet
Propulsion Laboratory, California Institute of Technology, under
a contract with NASA.

\end{acknowledgements}


\bibliographystyle{aa}
\bibliography{aa.bib}

\clearpage

\begin{table*}

\centering

\caption{Properties of Target Galaxies}
\label{table:gal_props}

\begin{tabular}{lccccc}
\hline\hline
Galaxy & $L_{H\alpha}$\tablefootmark{a} & SFR\tablefootmark{b} & Distance & Area Analyzed & Sources \\
Name   & (log ergs s$^{-1}$) & ($M_{\odot}$ yr$^{-1}$) & (mag)/(Mpc)    & (deg$^{2}$) & \\
\hline
NGC 55   & 40.55 & 0.28 & 26.58$\pm$0.11\tablefootmark{c} / 2.1$\pm$0.1 & 0.09  & 8746\\
NGC 253  & 40.99 & 0.78 & 27.70$\pm$0.07\tablefootmark{d} / 3.5$\pm$0.1 & 0.2   & 9001\\
NGC 2366 & 40.10 & 0.10 & 27.52\tablefootmark{e}          / 3.2         & 0.008 & 505\\
NGC 4214 & 40.19 & 0.12 & 27.10$\pm$0.18\tablefootmark{f} / 2.7$\pm$0.3 & 0.008 & 1185\\
NGC 5253 & 40.34 & 0.17 & 27.88$\pm$0.11\tablefootmark{g} / 3.6$\pm$0.2 & 0.009 & 722\\
\hline
\end{tabular}
\tablefoot{
\tablefoottext{a}{From \citet{ken08}}\\
\tablefoottext{b}{Derived from the H$\alpha$ luminosities in \citet{ken08}
  converted to star formation rates via equation 2 in \citet{ken98}}\\
\tablefoottext{c}{\citet{tan11}}\\
\tablefoottext{d}{\citet{rad11}}\\
\tablefoottext{e}{\citet{kar02}}\\
\tablefoottext{f}{\citet{dro02}}\\
\tablefoottext{g}{\citet{sak04}}\\
}
\end{table*}

\begin{table*}

\centering

\caption{Mid-infrared \textit{Spitzer} Photometry of 8746 Point Sources in the field of NGC 55}
\label{table:NGC55}

\begin{tabular}{ccrcrcrcrc}
\hline\hline
RA(J2000) & Dec(J2000) & [3.6] & $\sigma_{3.6}$ & [4.5] & $\sigma_{4.5}$ & [5.8] & $\sigma_{5.8}$ & [8.0] & $\sigma_{8.0}$ \\
  (deg)   &    (deg)   & (mag) & (mag)         & (mag) & (mag)         & (mag) & (mag)         & (mag) & (mag)  \\
\hline              	
     3.69063 & --39.24317 &  9.35 &  0.27 & 10.83 & 0.14 &  9.25 & 0.01 &  9.35 & 0.01 \\
     3.86827 & --39.26287 &  9.44 &  0.05 &  9.57 & 0.04 &  9.50 & 0.01 &  9.47 & 0.01 \\
     4.06058 & --39.19439 & 10.66 &  0.31 & 11.91 & 0.06 & 10.56 & 0.01 & 10.53 & 0.01 \\
     3.48295 & --39.16523 & 10.88 &  0.04 & 10.86 & 0.04 & 10.86 & 0.01 & 10.88 & 0.01 \\
     3.75037 & --39.21202 & 11.88 &  0.04 & 10.99 & 0.02 & 10.27 & 0.01 &  9.59 & 0.02 \\
     3.99595 & --39.17790 & 11.94 &  0.09 & 11.95 & 0.03 & 11.92 & 0.01 & 11.92 & 0.01 \\
     3.95731 & --39.28375 & 12.01 &  0.04 & 12.03 & 0.04 & 12.00 & 0.01 & 12.08 & 0.01 \\
     3.41756 & --39.12945 & 12.15 &  0.04 & 12.15 & 0.03 & 12.12 & 0.01 & 12.15 & 0.01 \\
     3.69610 & --39.22021 & 12.35 &  0.03 & 12.33 & 0.02 & 12.32 & 0.01 & 12.37 & 0.02 \\
     3.76566 & --39.13446 & 12.53 &  0.03 & 12.51 & 0.02 & 12.50 & 0.01 & 12.52 & 0.01 \\
\hline
\end{tabular}
\begin{center}
This table is available in its entirety in a machine-readable form
in the online journal. A portion is shown here for guidance 
regarding its form and content.
\end{center}
\end{table*}

\begin{table*}

\centering

\caption{Mid-infrared \textit{Spitzer} Photometry of 9001 Point Sources in the field of NGC 253}
\label{table:NGC253}

\begin{tabular}{ccrcrcrcrc}
\hline\hline
RA(J2000) & Dec(J2000) & [3.6] & $\sigma_{3.6}$ & [4.5] & $\sigma_{4.5}$ & [5.8] & $\sigma_{5.8}$ & [8.0] & $\sigma_{8.0}$ \\
  (deg)   &    (deg)   & (mag) & (mag)         & (mag) & (mag)         & (mag) & (mag)         & (mag) & (mag)  \\
\hline
    11.88921 & --25.28733 &  8.45 &  0.14 &  9.17 & 0.31 &  5.22 & 0.01 &  3.44 & 0.01 \\
    11.61232 & --25.38753 &  9.17 &  0.05 &  9.29 & 0.04 &  9.31 & 0.01 &  9.30 & 0.01 \\
    12.10072 & --25.41748 &  9.41 &  0.07 &  9.36 & 0.07 &  9.39 & 0.01 &  9.40 & 0.01 \\
    11.89076 & --25.28652 &  9.51 &  0.14 &  9.72 & 0.20 &  6.61 & 0.01 &  4.79 & 0.01 \\
    11.72243 & --25.24591 &  9.58 &  0.05 &  9.64 & 0.04 &  9.61 & 0.01 &  9.63 & 0.01 \\
    11.96337 & --24.95917 &  9.61 &  0.05 &  9.61 & 0.08 &  9.69 & 0.01 &  9.75 & 0.01 \\
    12.01735 & --25.07951 &  9.97 &  0.09 &  9.73 & 0.10 &  8.58 & 0.01 &  8.52 & 0.01 \\
    11.63976 & --25.45913 & 10.08 &  0.06 & 10.12 & 0.03 & 10.12 & 0.01 & 10.11 & 0.01 \\
    12.13264 & --25.22499 & 10.53 &  0.03 & 10.57 & 0.03 & 10.52 & 0.01 & 10.54 & 0.01 \\
    11.85995 & --25.32108 & 10.59 &  0.03 & 10.60 & 0.03 & 10.54 & 0.01 & 10.45 & 0.02 \\
\hline
\end{tabular}
\begin{center}
This table is available in its entirety in a machine-readable form
in the online journal. A portion is shown here for guidance 
regarding its form and content.
\end{center}
\end{table*}

\begin{table*}

\centering

\caption{Mid-infrared \textit{Spitzer} Photometry of 505 Point Sources in the field of NGC 2366}
\label{table:NGC2366}

\begin{tabular}{ccrcrcrcrc}
\hline\hline
RA(J2000) & Dec(J2000) & [3.6] & $\sigma_{3.6}$ & [4.5] & $\sigma_{4.5}$ & [5.8] & $\sigma_{5.8}$ & [8.0] & $\sigma_{8.0}$ \\
  (deg)   &    (deg)   & (mag) & (mag)         & (mag) & (mag)         & (mag) & (mag)         & (mag) & (mag)  \\
\hline
   112.12196 &  69.23869 & 12.07 &  0.09 & 11.88 &  0.11 &  12.11 & 0.01 & 12.07 & 0.01 \\
   112.18302 &  69.18944 & 13.23 &  0.08 & 12.30 &  0.05 &  11.29 & 0.01 &  9.98 & 0.01 \\
   112.23172 &  69.21793 & 13.26 &  0.03 & 13.09 &  0.06 &  13.10 & 0.01 & 13.06 & 0.01 \\
   112.18950 &  69.20526 & 13.49 &  0.13 & 13.43 &  0.10 &  12.91 & 0.01 & 10.21 & 0.01 \\
   112.14255 &  69.20529 & 14.01 &  0.06 & 14.05 &  0.03 &  14.03 & 0.01 & 14.12 & 0.02 \\
   112.13716 &  69.22132 & 14.27 &  0.06 & 14.30 &  0.04 &  14.30 & 0.02 & 14.40 & 0.04 \\
   112.36899 &  69.20294 & 14.52 &  0.06 & 14.39 &  0.10 &  14.43 & 0.02 & 14.11 & 0.05 \\
   112.21034 &  69.20583 & 14.53 &  0.05 & 14.58 &  0.04 &  15.29 & 0.04 & 14.48 & 0.04 \\
   112.05402 &  69.21157 & 14.60 &  0.10 & 14.59 &  0.07 &  14.51 & 0.03 & 15.20 & 0.09 \\
   112.04636 &  69.20489 & 14.60 &  0.09 & 14.05 &  0.08 &  12.16 & 0.01 & 12.24 & 0.01 \\
\hline
\end{tabular}
\begin{center}
This table is available in its entirety in a machine-readable form
in the online journal. A portion is shown here for guidance 
regarding its form and content.
\end{center}
\end{table*}

\begin{table*}

\centering

\caption{Mid-infrared \textit{Spitzer} Photometry of 1185 Point Sources in the field of NGC 4214}
\label{table:NGC4214}

\begin{tabular}{ccrcrcrcrc}
\hline\hline
RA(J2000) & Dec(J2000) & [3.6] & $\sigma_{3.6}$ & [4.5] & $\sigma_{4.5}$ & [5.8] & $\sigma_{5.8}$ & [8.0] & $\sigma_{8.0}$ \\
  (deg)   &    (deg)   & (mag) & (mag)         & (mag) & (mag)         & (mag) & (mag)         & (mag) & (mag)  \\
\hline
   183.89776 & 36.36563 & 12.11 & 0.06 & 12.21 & 0.06 & 12.24 & 0.01 & 12.21 & 0.01\\
   183.98066 & 36.32288 & 12.43 & 0.10 & 12.62 & 0.07 & 12.63 & 0.01 & 12.57 & 0.01\\
   183.92065 & 36.31775 & 12.95 & 0.14 & 12.43 & 0.10 & 10.22 & 0.00 &  8.45 & 0.00\\
   183.94470 & 36.35236 & 12.99 & 0.05 & 13.00 & 0.03 & 13.04 & 0.01 & 13.04 & 0.01\\
   183.90811 & 36.37277 & 13.22 & 0.09 & 13.09 & 0.10 & 13.23 & 0.01 & 13.00 & 0.02\\
   183.91366 & 36.32601 & 13.25 & 0.12 & 12.75 & 0.10 & 10.62 & 0.01 &  9.03 & 0.02\\
   183.90514 & 36.37191 & 13.34 & 0.06 & 13.34 & 0.05 & 13.29 & 0.01 & 13.35 & 0.04\\
   183.91279 & 36.32693 & 13.36 & 0.10 & 12.81 & 0.10 & 10.82 & 0.01 &  9.08 & 0.01\\
   183.90902 & 36.32899 & 13.50 & 0.05 & 13.65 & 0.06 & 13.63 & 0.13 & 12.91 & 0.30\\
   183.91710 & 36.32505 & 13.74 & 0.15 & 13.53 & 0.14 & 11.02 & 0.01 &  9.30 & 0.01\\
\hline
\end{tabular}
\begin{center}
This table is available in its entirety in a machine-readable form
in the online journal. A portion is shown here for guidance 
regarding its form and content.
\end{center}
\end{table*}

\begin{table*}

\centering

\caption{Mid-infrared \textit{Spitzer} Photometry of 722 Point Sources in the field of NGC 5253}
\label{table:NGC5253}

\begin{tabular}{ccrcrcrcrc}
\hline\hline
RA(J2000) & Dec(J2000) & [3.6] & $\sigma_{3.6}$ & [4.5] & $\sigma_{4.5}$ & [5.8] & $\sigma_{5.8}$ & [8.0] & $\sigma_{8.0}$ \\
  (deg)   &    (deg)   & (mag) & (mag)         & (mag) & (mag)         & (mag) & (mag)         & (mag) & (mag)  \\
\hline
   204.98313 & --31.64007 &  9.27 & 0.06 &  7.99 & 0.04 &  6.30 & 0.01 &  4.78 & 0.01\\
   204.92952 & --31.63631 & 10.49 & 0.04 & 10.50 & 0.02 & 10.50 & 0.01 & 10.52 & 0.01\\
   204.97971 & --31.61547 & 10.67 & 0.06 & 10.81 & 0.03 & 10.73 & 0.01 & 10.80 & 0.01\\
   205.02464 & --31.61900 & 11.56 & 0.05 & 11.67 & 0.04 & 11.57 & 0.01 & 11.68 & 0.01\\
   204.93852 & --31.60614 & 11.74 & 0.05 & 11.78 & 0.05 & 11.66 & 0.01 & 11.85 & 0.01\\
   204.95026 & --31.62672 & 12.64 & 0.04 & 12.74 & 0.04 & 12.64 & 0.01 & 12.82 & 0.02\\
   204.96348 & --31.61572 & 13.18 & 0.02 & 13.24 & 0.02 & 13.18 & 0.01 & 13.41 & 0.02\\
   204.98281 & --31.64374 & 13.45 & 0.23 & 13.29 & 0.16 & 11.17 & 0.04 &  9.66 & 0.05\\
   204.98523 & --31.63999 & 13.63 & 0.26 & 13.59 & 0.33 &  9.40 & 0.01 &  5.12 & 0.01\\
   204.95859 & --31.64485 & 13.87 & 0.03 & 13.91 & 0.03 & 14.19 & 0.02 & 13.95 & 0.03\\
\hline
\end{tabular}
\begin{center}
This table is available in its entirety in a machine-readable form
in the online journal. A portion is shown here for guidance 
regarding its form and content.
\end{center}
\end{table*}

\clearpage

\clearpage

\clearpage

\clearpage

\clearpage

\clearpage

\clearpage

\clearpage

\clearpage

\clearpage

\clearpage

\clearpage

\clearpage

\clearpage

\end{document}